\documentclass{article}

\usepackage{amssymb}
\usepackage{amsmath}
\usepackage{supertabular}
\usepackage{longtable}
\usepackage{array}
\usepackage{amstext}
\usepackage{epsfig}
\begin{document}
\title{Asymptotic states in brane cosmology with a
nonlocal anisotropic stress}
\author{N. Yu. Savchenko$^{1}$, \ S.V. Savchenko$^{2}$ \
   and \ A. V. Toporensky$^{1}$}
\date{}
\maketitle
\hspace{-6mm}
$^{1}${\em Sternberg
Astronomical Institute, Moscow University, Moscow 119899, Russia}\\
$^{2}${\em Landau Institute for Theoretical Physics,
Russian Academy of Sciences, Kosygin Street 2, Moscow 117334,
Russia}

\begin{abstract}
We investigate the dynamics of a Bianchi I brane Universe in the presence
of a nonlocal anisotropic stress ${\cal P}_{\mu\nu}$
proportional to a "dark energy" ${\cal U}$. Using this ansatz for the
case ${\cal U} > 0$ we prove that if a matter on a brane satisfies
the equation of state $p=(\gamma-1)\rho$ with $\gamma \le 4/3$
then all such models isotropize. For $\gamma > 4/3$ anisotropic
future asymptotic states are found. We also describe the past asymptotic
regimes for this model.
\end{abstract}

\section{Introduction}

The idea that our space has more than three dimensions was attractive
during many years, mostly in the scenario of compactification.
Recently a new paradigm appeared. In this new approach
the matter is trapped in three-dimensional brane embedded in a
multi-dimensional space (bulk)~\cite{R-S}. In the case of one extra 
dimension it is
shown that
isotropic cosmological evolution being unusual in early stages of
the brane Universe~\cite{B-L-D} tends to the standard cosmological
scenario in the late time~\cite{B-L-D2}.

The field equations on the brane are~\cite{Maeda,Maartens}
\begin{equation}
G_{\mu \nu}=\kappa^2 T_{\mu \nu} +
\tilde \kappa^4 S_{\mu \nu} - {\cal E}_{\mu \nu}.
\end{equation}
Here $\tilde \kappa$ is the fundamental 5-dimensional gravitational constant,
$\kappa$ is the effective 4-dimensional gravitation constant on the brane.
Bulk corrections to the Einstein equations on the brane are of two forms:
there are quadratic energy-momentum corrections via the tensor $S_{\mu \nu}$
and nonlocal effects from the free gravitational field in the bulk
${\cal E}_{\mu\nu}$. The quadratic corrections are significant only in
the early stages of cosmological evolution (see the next section).
 The nonlocal
corrections can be decomposed into a scalar (nonlocal energy density),
a vector (nonlocal energy flux) and a tensor (nonlocal anisotropic stress)
parts~\cite{Maartens}. On an isotropic brane there is only a scalar part $\cal U$ which
decays during the expansion of the Universe as ${\cal U} \sim a^{-4}$ where
$a$ is the scale factor.

However, if we remove the suggestion of isotropy  (for example, 
in order to study
the problem of isotropization of the brane Universe), then the tensor
part of nonlocal contribution ${\cal P}_{\mu\nu}$ should be taken into
account. The main obstacle to do this is the absence of an evolution
equation for the anisotropic stress ${\cal P}_{\mu\nu}$~\cite{Maartens,Varun}.
Its correct form should be derived from a full $5$-dimensional anisotropic
metric which is still unknown. In this situation we have to choose some
ansatz for ${\cal P}_{\mu\nu}$. In first papers on Bianchi I brane either
all nonlocal corrections~\cite{Varun,C-S1} or its tensor part~\cite{C-S2,Topor,Hervik}
have been neglected. This approach gives some interesting results such as
isotropic character of the initial singularity~\cite{Varun,C-S1} and the possible
existence of an intermediate anisotropic stage~\cite{Topor}. However, whether
these results remain valid when the contribution from ${\cal P}_{\mu\nu}$
becomes important is still an open problem.

Recently Barrow and Maartens have proposed an ansatz for the nonlocal stress
which generalizes known anisotropic sources in General Relativity~\cite{B-M}.
They have studied the dynamics of isotropization in the limit of small
anisotropy. In the present paper we use this ansatz for describing
the global properties of the dynamics. We use the techniques of dynamical
systems and the expansion normalized variables (for the description of
this formalism see~\cite{Ellis}) to find the past and the future attractors of
our system. The transient behavior is also covered by studying the future
asymptotic states for a pure nonstandard (i.e. with $S_{\mu\nu}$ and without $T_{\mu\nu}$
terms in (1)) brane dynamics. Of course, in this framework we can not find
regimes originated from an interplay between standard (proportional to
$T_{\mu\nu}$) and nonstandard  (proportional to $S_{\mu\nu}$) source terms.
Up to now several interesting regimes of this type have been found (for example,
an analog of
the Einstein static Universe~\cite{Ge} and a stable oscillationary regime~\cite{C-S2}).

The paper is organized as follows: in Sec.2. we describe the dimensionless
variables and use them to write down the system of differential equations
in a convenient form. In Sec.3 we list stable points of this system neglecting
the square energy-momentum correction and describe the
future asymptotic states of the Bianchi I brane model. In Sec.4
the stable points of the system without linear energy-momentum contribution
are listed and possible transient behavior is obtained.  We also discuss
the past asymptotic states of our model. Sec.5 provides a brief summary of the results
obtained.

\section{Expansion normalized variables and equation of motion}

We will use an orthonormal frame of vector fields ${\it e}_a$ such that
${\it e}_0=\partial_{t}$. In the Bianchi I case we have the following commutation
relations for the vector basis (Latin indices run from 0 to 3 and
Greek indices from 1 to 3): 

$$
[{\it e}_a,{\it e}_b]=C^{c}_{ab} {\it e}_c
$$
where

$$
C^{\mu}_{0\nu}=-\Theta^{\mu}_{\nu}-
\epsilon^{\mu}_{\nu \alpha} \Omega^{\alpha}, \qquad
C^{\gamma}_{\alpha \beta}=0.
$$

Here $\Omega^{\alpha}$ is the angular velocity of the spatial triad with respect
to a Fermi-propagated frame and $\Theta_{\mu \nu}$ can be decomposed into
a volume expansion rate $\Theta$ and a shear $\sigma_{\mu \nu}$:

$$
\Theta_{\mu \nu}=\frac{1}{3}\Theta \delta_{\mu \nu} +
\sigma_{\mu \nu}.
$$

If we introduce the metric for the Bianchi I model in the standard form
\begin{equation}
ds^2 = -dt^2 + a_i^2(t) (dx^i)^2,
\end{equation}
then the volume expansion rate in terms of the main scale factor
$a=(a_1 a_2 a_3)^{1/3}$ is $\Theta = 3H = 3\frac{\dot a}{a}$, where
$H$ is the main Hubble parameter.

 For these variables we can write
the Raychaudhuri equation
\begin{equation}
\dot\Theta+\frac13\Theta^2+\sigma^{\mu\nu}\sigma_{\mu\nu}+\frac12\kappa^2
(\rho+3p)=-\frac12\kappa^2(2\rho+3p)\frac{\rho}{\lambda}-\frac{6\cal
U}{\kappa^2\lambda},
\end{equation}
the Gauss-Codazzi equations
\begin{eqnarray}
\dot{\sigma}_{\mu\nu}+\Theta\sigma_{\mu\nu}&=&{6
\over\kappa^2\lambda}{\cal P}_{\mu\nu},\\
-\frac{2}{3}\Theta^2 +\sigma^{\mu\nu}\sigma_{\mu\nu}
+2\kappa^2\rho &=& -\kappa^2{\rho^2\over\lambda} -{12 {\cal
U}\over\kappa^2\lambda}
\end{eqnarray}
and  the conservation
equations
\begin{eqnarray}
&&\dot{\rho}+\Theta(\rho+p)=0,\\ && \dot{\cal
U}+\frac{4}{3}\Theta{\cal U}+\sigma^{\mu\nu}{\cal
P}_{\mu\nu}  =0 , \\&& D^{\nu}{\cal P}_{\mu\nu}
=0.
\end{eqnarray}
Derivation of these equations from Eq.(1) and the explicit form of
$S_{\mu\nu}$ see in~\cite{Maartens}.
Here $\lambda$ is the brane tension. Quadratic energy-momentum corrections
(first terms in RHS of Eqs. (3) and (5))
become significant if a matter density $\rho$ on the brane is greater
than $\lambda$. The current experimental limit on the brane tension is
$\lambda > 100(GeV)^4$~\cite{limit}.

Due to presence of the nonlocal stress ${\cal P}_{\mu \nu}$ this system
of equations is not closed. Therefore, it is necessary to specify the
form of ${\cal P}_{\mu \nu}$ using some additional suggestion. In the
present paper we deal with the proposal of Barrow and Maartens~\cite{B-M}.
They have suggested that the nonlocal stress is proportional to the nonlocal
energy density
\begin{equation}
{\cal P}_{\mu\nu}=2D_{\mu\nu}{\cal U},
\end{equation}
\begin{equation*}
D_{\mu\nu}=\mathrm{diag}\{c_1,c_2,-(c_1+c_2)\},
\end{equation*}
where $c_1$, $c_2$~ are some constant.

It is worth to notice that choosing the appropriate coordinate system we
can diagonalize either ${\cal P}_{\mu \nu}$ or $\sigma_{\mu \nu}$. We
choose to make ${\cal P}_{\mu \nu}$ diagonal.

Nevertheless, our system of equations is not completely
determined yet. We must fix the angular velocity $\Omega^{\alpha}$ of the spatial
triad associated with ${\cal P}_{\mu \nu}$. After that we can write the
derivatives of the shear tensor as

$$
\dot\sigma_{\mu\nu}=\partial_{t}\sigma_{\mu\nu}+
2\sigma^{\delta}_{(\mu}\epsilon_{\nu)\gamma\delta}\Omega^{\gamma}.
$$

In the GR description of cosmological anisotropic sources (for example,
the homogeneous magnetic field~\cite{LeBlanc}) the angular velocity
components are constrained by the corresponding equations of motion
of the anisotropic fluid
(for example, the Maxwell equations in the magnetic field case).
  As we have no equation
for ${\cal P}_{\mu \nu}$, we should choose an ansatz for $\Omega^{\alpha}$.
In the present paper we consider the simplest possible form 
$\Omega^{\alpha}=0$
(in the end of the next section one important statement not depending
on $\Omega^{\alpha}$ is proved).

After introduction of the deceleration parameter
\begin{equation}
q=\displaystyle\frac{\ddot a a}{\dot a^2}
\end{equation}
and dimensionless variables
\begin{align}
\Sigma_+&=\displaystyle\frac32(\sigma_{22}+\sigma_{33})/\Theta,\notag
\lower 3mm\hbox{}\\
\Sigma_-&=\displaystyle\frac12\sqrt3(\sigma_{22}-\sigma_{33})/\Theta,\notag
\lower 3mm\hbox{}\\
\Sigma_{12}&=\sqrt3\sigma_{12}/\Theta,\lower 3mm\hbox{}\\
\Sigma_{23}&=\sqrt3\sigma_{23}/\Theta,\notag\lower 3mm\hbox{}\\
\Sigma_{13}&=\sqrt3\sigma_{13}/\Theta,\notag\lower 3mm\hbox{}\\
U&=\displaystyle\frac{18{\cal U}}{\kappa^2\lambda \Theta^2},\notag
\end{align}
we may rewrite our system in a simpler form.

We have also auxiliary evolution equation for the matter density.
There are two ways for description of the quadratic energy-momentum terms.
It is possible either to incorporate them into a single evolution equation for
the matter as it have been done in~\cite{Hervik} or to consider separately as some
kind of a particular effective matter~\cite{C-S1,C-S2}.
We choose the latter case and write
two evolution equations - one for "standard" matter and the
other for "nonstandard" matter.
In order to do this we introduce dimensionless variables
\begin{equation}
\begin{array}{rl}
\label{npo}
\Omega_\lambda&=\displaystyle\frac{3\kappa^2\rho^2}{2\lambda \Theta^2},
\lower 5mm\hbox{}\\
\Omega_\mu&=\displaystyle\frac{3\kappa^2\rho}{\Theta^2}.\lower 5mm\hbox{}\\
\end{array}
\end{equation}
Note that these values are not independent, they connected to each other by
$\Omega_{\lambda}/\Omega_{\mu}=\rho/2\lambda$.

Finally, we introduce a new time variable $\tau$ such that
$\frac{dt}{d\tau}=\frac{3}{\Theta}$.
Now our system becomes (prime denotes a derivative with respect to $\tau$)
\begin{align}
\label{si}
U'&=2U[(q-1)-3\Sigma_+(c_1+c_2)-\sqrt3\Sigma_-(c_1-c_2)],\notag
\lower 3mm\hbox{}\\
\Sigma_+'&=3U(c_1+c_2)+
\Sigma_+(q-2),\notag\lower 3mm\hbox{}\\
\Sigma_-'&=\sqrt3 U (c_1-c_2)
+\Sigma_-(q-2),\lower 3mm\hbox{}\\
\Sigma_{12}'&=(q-2)\Sigma_{12}
,\notag\lower 3mm\hbox{}\\
\Sigma_{13}'&=(q-2)\Sigma_{13}
,\notag\lower 3mm\hbox{}\\
\Sigma_{23}'&=(q-2)\Sigma_{23} \notag
\end{align}
with the constraint
\begin{equation}
1=\Sigma^2+\Omega_{\lambda}+\Omega_{\mu}+U,
\end{equation}
the equation for $q$
\begin{equation}
\label{Qq}
q=2\Sigma^2+\displaystyle\frac12(3\gamma-2)\Omega_{\mu}+(3\gamma-1)\Omega_{
\lambda}+U,
\end{equation}
and the evolution equations for matter density
\begin{align}
\Omega_{\mu}'&=[2(1+q)-3\gamma]\Omega_{\mu},\lower 3mm\hbox{}\\
\Omega_{\lambda}'&=[2(1+q)-6\gamma]\Omega_{\lambda}.
\end{align}
Here
\begin{equation}
\Sigma^2=\Sigma_+^2+\Sigma_-^2+\Sigma_{12}^2+\Sigma_{23}^2+\Sigma_{13}^2.
\end{equation}

There is an important difference between GR and brane problems. In common
GR formulation of the problem the energy density of an anisotropic matter
is positive, though the nonlocal energy density on the brane can be either
positive or negative and, in principal, can change sign in the cosmological
evolution. However, ansatz (9), as we can see from the first
equation of system (13) prevents
${\cal U}$ from the change of the sign. The case of negative energy of an
anisotropic matter has not been investigated previously in the GR formulation
due to its peculiarity from the physical point of view,
but in the brane scenario it becomes important. The main feature which
arises from ${\cal U} < 0$ is that the flat Bianchi I Universe can
recollapse
without reaching any future nonsingular asymptotic regime.
From the mathematical point of view the negativity of ${\cal U}$ makes
our phase space noncompact (the phase space, specially constracted
to be compact in the ${\cal U} < 0$ case see in \cite{C-S2}).
So, from this point we
assume ${\cal U} >0$ leaving the opposite case to a future work. Now our phase
space becomes compact with $\Sigma_+, \Sigma_-, \Sigma_{12}, \Sigma_{13},
\Sigma_{23}$ varying in the range $[-1,1]$ and $\Omega_{\mu},
\Omega_{\lambda}, U$ varying in the range $[0,1]$.

In the next section we start to analyze the system (13) -- (17) neglecting
the "nonstandard" matter.
All the future
attractors for this system are also the future attractors for the initial
general problem, because for $t \to \infty$ we can neglect a nonstandard
matter on the brane in comparison with a standard one.

\section{Equilibrium points and future evolution}

Now we list equilibrium points for the system (13) -- (17)
in the case $\Omega_{\lambda}=0$.
It is convenient to use the notations $c_{+} = c_1+c_2$ and $c_{-} = c_1-c_2$.
We also have found the corresponding eigenvalues $\lambda_i$
of the linearization of system (13) about the equilibrium points
and have used
the condition $\lambda_i < 0$ to study the stability of the equilibrium
points in the future direction.
\begin{enumerate}
\item  The isotropic Universe
\begin{align*}
&U=\Sigma_+=\Sigma_-=\Sigma_{12}=\Sigma_{13}=\Sigma_{23}=0\lower 5mm\hbox{}\\
&q=\displaystyle\frac32\gamma-1 \qquad\Omega_{\mu}=1
\qquad 0\leqslant\gamma\leqslant2\hbox to 10cm {}
\end{align*}
Stable for $\gamma < 4/3$ independently on $c_+ , c_-$.
\item Equilibrium point
\begin{align*}
&U=\frac{(2-\gamma)(3\gamma-4)}{12c_{+}^2+4c_{-}^2}\qquad
\Sigma_+=\displaystyle\frac{c_{+}(3\gamma-4)}{6c_{+}^2+2c_{-}^2}
\lower 5mm\hbox{}\\
&\Sigma_-=\frac{\sqrt{3}c_{-}(3\gamma-4)}{18 c_{+}^2+6c_{-}^2}\qquad
\Omega_{\mu}=1-\frac{3\gamma-4}{18c_{+}^2+6c_{-}^2}
\lower 5mm\hbox{}\\
&\Sigma_{12}=\Sigma_{13}=\Sigma_{23}=0 \lower 5mm\hbox{}\\
&\gamma \in (4/3, 2)\hbox to 10cm {}
\end{align*}
Stable for
$\gamma<4/3+6c_+^2+2c_-^2$.
\item Equilibrium point
\begin{align*}
&U=1-9c_+^2-3c_-^2 \qquad \Sigma_+=3c_+ \qquad \Sigma_-=\sqrt{3}c_-
\lower 5mm\hbox{}\\
&\Sigma_{12}=\Sigma_{13}=\Sigma_{23}=0 \lower 5mm\hbox{}\\
&\Omega_{\mu}=0 \qquad 3c_+^2 + c_-^2 <1/3
\hbox to 10cm {}
\end{align*}
Stable for $\gamma > 4/3 + 6c_+^2 + 2c_-^2$.
\item Equilibrium set (the Kasner Universe)
\begin{align*}
&U=0 \qquad \Omega_{\mu}=0 \qquad q=2 \lower 5mm\hbox{}\\
&\Sigma_+^2 + \Sigma_-^2 + \Sigma_{12}^2 + \Sigma_{13}^2 + \Sigma_{23}^2=1 \hbox to 10cm {}
\end{align*}
\end{enumerate}

The results on stability of the equilibrium points 
are presented graphically in Fig.1. The space
$(\gamma, c_+, c_-)$ is divided into several zones in which only one stable
future attractor exists.

\begin{figure}
\epsfxsize=10cm
\centerline{\epsfbox{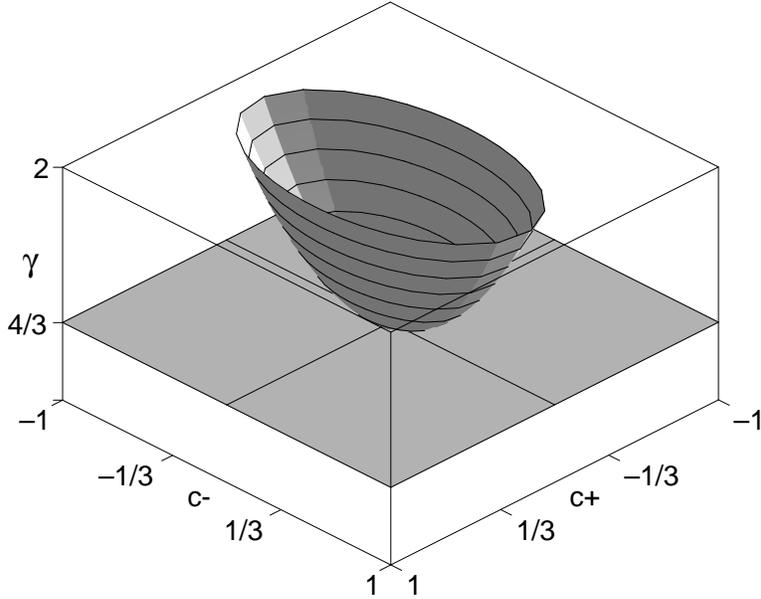}}
\caption{Stability zones of the equilibrium points for the case $\Omega_
{\lambda}=0$. Below the plane the point 1 is stable, above
the plane outside the compact
region the point 2 is stable, inside the compact region the point 3
is stable.}
\end{figure}

We can see that the only stable point for $\gamma < 4/3$ is the isotropic
one, independently on $c_+$ and $c_-$. It is possible to strengthen this result
and prove that all the solutions with $\Omega_{\mu} > 0$ and $\gamma \le 4/3$
tend to the isotropic solution ($\Omega_{\mu}=0$ is an invariant subset
of our phase space, as it can be seen from Eq. (16)). We point out that a physically
important boundary case $\gamma = 4/3$ is included in the conditions of this
statement. The key point is the equation
\begin{equation}
\Omega_{\mu}^{\prime}=[(6-3\gamma)\Sigma^{2}+
(4-3\gamma)U]\Omega_{\mu}
\end{equation}
which can be obtained from Eqs. (14) -- (16).
Note, that Eq.(19) follows only from the Raychaudhuri and constraint equations
and we do not use the evolution equations for the shear. This means that
the consideration presented below does not depend on a particular choice
of the angular velocity $\Omega^{\alpha}$.

First, consider the case $\gamma < 4/3.$
Then there is a positive number $\alpha$ such that
$$\Omega_{\mu}^{\prime}\ge
\alpha[\Sigma^{2}+U]\Omega_{\mu}.$$
From this inequality it follows that
the quantity
$\Omega_{\mu}$ is strictly increasing if $\Omega_{\mu}(\tau_{0})> 0.$
Using this fact and the constraint
$\Sigma^{2}+\Omega_{\mu}+U=1,$
one can easily understand
that the sum $\Sigma^{2}+U,$ in turn, strictly decreases.
 As decreasing and bounded below by zero,
this sum has some limit $\beta \ge 0$.
Assume that $\beta$ is not equal to zero.
Then,  the quantity
$\Omega_{\mu}$ goes to infinity:
$\ln \Omega_{\mu}(\tau)-\ln \Omega_{\mu}(\tau_{0})>
 \alpha\beta(\tau-\tau_{0})$ which
contradicts the inequality $\Omega_{\mu}(\tau) \le 1$.
Thus, for any initial conditions satisfying
$ \Omega_{\mu}(\tau_{0}) > 0$
the solution tends to the point
$\Sigma_{+}=\Sigma_{-}=\Sigma_{12}=\Sigma_{13}=\Sigma_{23}=U=0.$

In the boundary case $\gamma = 4/3$ it is necessary to be more careful.
For $\gamma=4/3$ we have
\begin{equation}
\Omega_{\mu}^{\prime}=2\Sigma^{2}\Omega_{\mu}.
\end{equation}
So, $\Omega_{\mu}$ is monotone increasing and, then, tends to some
limit value $\Omega_{lim}$. Assume that $\Omega_{lim} < 1$.
 Using the constraint (14) and the evolution equations
for $U$ and $\Omega_{\mu}$, we obtain the evolution equation for
$\Sigma^2$:
$$\frac{\partial}{\partial \tau}\Sigma^{2}=6(c_{1}+c_{2})U\Sigma_{+}+
2\sqrt{3}(c_{1}-c_{2})U\Sigma_{-}+2\Sigma^{2}(\Sigma^{2}-1).$$
In sequel, we shell use only the fact that
the absolute value of the LHS of this equation is bounded
above:
$|\frac{\partial}{\partial \tau}\Sigma^{2}(\tau)|< C$ for all $\tau$.
Now assume that $\Sigma^{2}$ does not tend to zero.
Then there is a sequence
$\tau_{n}$
going to infinity
such that
$\Sigma^{2}(\tau_{n})> c$  for some $c > 0.$
But the boundedness of the shear square derivative
means that
there is a positive $b$ which does not depend on $n$
such that
$\Sigma^{2}(\tau)> c/2$ for $\tau\in (\tau_{n}-b,\tau_{n}+b)$
and all $n.$
In this case from equation (20) it follows that the quantity
$\Omega_{\mu}$ goes to infinity:
$$\ln \Omega_{\mu}(\tau)-\ln \Omega_{\mu}(\tau_{0})=\int\limits_{\tau_{0}}^{\tau}
2\Sigma^{2}dt > \sum\limits_{n:\tau_{n}< \tau}2cb.$$
This fact contradicts the inequality
$\Omega_{\mu}(\tau) \le 1$.
Thus, $\Sigma^{2}$ tends to zero.
By the constraint
$1=\Sigma^{2}+\Omega_{\mu}+U,$ this means that
the quantity $U$ converges to $1-\Omega_{lim}>0.$
But it is impossible because
the point $\Sigma^{2}=0$ and $U=const > 0$
is not an equilibrium point of our system
for nonzero $c_1$ and $c_2$.
Thus, $\Omega_{\mu}$ goes to $1$ and both $U$ and $\Sigma^{2}$
decrease to zero.

For the brane problem this means that if $\cal P_{\mu\nu}$ satisfies
the ansatz (9) with ${\cal U} >0$, then an arbitrary
anisotropy of the whole 5-dimensional metric does not destroy the
future isotropic regime unless $\gamma>4/3$.

It is interesting that in the case of $\gamma=4/3$ without an
anisotropic stress we have a one-dimensional set of equilibria instead
of the isotropic equilibrium point. Points of this set have $\Sigma^2=0$ and
are marked by the constant ratio
$U/\Omega_{\mu}$. A nonzero ${\cal P}_{\mu\nu}$ makes all these
points except for $U=0, \Omega_{\mu}=1$ unstable.
As a result, in the presense of a nonlocal anisotropic stress
the "dark energy" ${\cal U}$ can be washed out
during the cosmological evolution even in the radiation-dominated epoch.

We finish this section by one remark which is not directly connected
with the brane cosmology. As we consider the case of $\Omega_{\lambda}=0$, the
system of equations (13) -- (16) with Barrow-Maartens ansatz for ${\cal P}_{\mu\nu}$ has the same
form (with appropriate rescaling of source terms) as the system describing
the evolution of the Universe in the presence of an anisotropic matter
in the ordinary General Relativity. Since our result on isotropisation
does not depend on angular velocity $\Omega^{\alpha}$ (these values should
be determined separately for each particular anisotropic source),
we can claim that in the presence of
an anisotropic matter with an anisotropic stress $\pi_{\mu\nu}$ proportional to
its energy density $\epsilon$ and an average pressure $\bar p$
equals to $\epsilon/3$ all the models isotropize if $\gamma \le 4/3$.
Moreover, we can consider an anisotropic matter with an average
pressure $\bar p=(\gamma_{eff}-1)\epsilon$ and, using the same procedure,
claim
that all such models with $\gamma \le \gamma_{eff}$ tend to the isotropic
attractor. This fact was previously discovered in the particular case
of the homogeneous magnetic field \cite{LeBlanc} and for the general source in
the limit of a small anisotropy \cite{Barrow1} (the "criticality condition"
of Ref. \cite{Barrow1} is exactly $\gamma = \gamma_{eff}$ in our
notation).

\section{A pure brane regime}

In this section we neglecte the standard matter on a brane.
This case describes a purely nonstandard brane dynamics when the matter density
on the brane is large in comparison with the brane tension. The past asymptotic
states
of these models correspond to a brane dynamics near a singularity, future
asymptotic states describe some intermediate regime of the brane Universe
when a nonstandard asymptotic behavior have already been established but the
matter density is still larger than the brane tension.

\begin{description}
\item[$1'$.] The isotropic Universe
\begin{align*}
&U=\Sigma_+=\Sigma_-=\Sigma_{12}=\Sigma_{13}=\Sigma_{23}=0\lower 5mm\hbox{}\\
&q=3\gamma-1 \qquad\Omega_{\lambda}=1
\qquad 0\leqslant\gamma\leqslant2\hbox to 10cm {}
\end{align*}
Stable in the future for $\gamma < 2/3$, stable in the past
for $\gamma > 1$.
\item[$2'$.] Equilibrium point
\begin{align*}
&U=\frac{(1-\gamma)(3\gamma-2)}{3c_{+}^2+c_{-}^2}\qquad
\Sigma_+=\displaystyle\frac{c_{+}(3\gamma-2)}{3c_{+}^2+c_{-}^2}
\lower 5mm\hbox{}\\
&\Sigma_-=\frac{\sqrt{3}c_{-}(3\gamma-2)}{9 c_{+}^2+3c_{-}^2}\qquad
\Omega_{\lambda}=1-\frac{3\gamma-2}{9c_{+}^2+3c_{-}^2}
\lower 5mm\hbox{}\\
&\Sigma_{12}=\Sigma_{13}=\Sigma_{23}=0 \lower 5mm\hbox{}\\
&\gamma \in (2/3, 1)\hbox to 10cm {}
\end{align*}
Stable in the future for
$\gamma<2/3+3c_+^2+c_-^2$.
\item[$3'$.] Equilibrium point
\begin{align*}
&U=1-9c_+^2-3c_-^2 \qquad \Sigma_+=3c_+ \qquad \Sigma_-=\sqrt{3}c_-
\lower 5mm\hbox{}\\
&\Sigma_{12}=\Sigma_{13}=\Sigma_{23}=0 \lower 5mm\hbox{}\\
&\Omega_{\lambda}=0 \qquad 3c_+^2 + c_-^2 <1/3
\hbox to 10cm {}
\end{align*}
Stable in the future for $\gamma > 2/3 + 3c_+^2 + c_-^2$.
\item[$4'$.] Equilibrium set (the Kasner Universe)
\begin{align*}
&U=0 \qquad \Omega_{\lambda}=0 \qquad q=2 \lower 5mm\hbox{}\\
&\Sigma_+^2 + \Sigma_-^2 + \Sigma_{12}^2 + \Sigma_{13}^2 + \Sigma_{23}^2=1 \hbox to 10cm {}
\end{align*}
Stable in the future for $3c_+^2 + c_-^2 > 1/3$.
\end{description}

Zones of stability of these equilibrium points are plotted in Fig.2.
For $\gamma<1$ we can see a complete analogy
with Fig.1 if $\gamma$  is divided by 2. It is natural, because
the nonstandard matter behaves effectively as a "normal"
matter with $\gamma_{NS}=2
\gamma$. Then, to receive expressions for equilibrium points and
stability zones of the system (13) -- (17) with $\Omega_{\mu}=0$
it is enough to take the corresponding expressions
for $\Omega_{\lambda}=0$ and replace $\gamma$ by $2\gamma$.
In particular, the isotropic attractor is stable
for $\gamma < 2/3$.

\begin{figure}
\epsfxsize=13cm
\centerline{\epsfbox{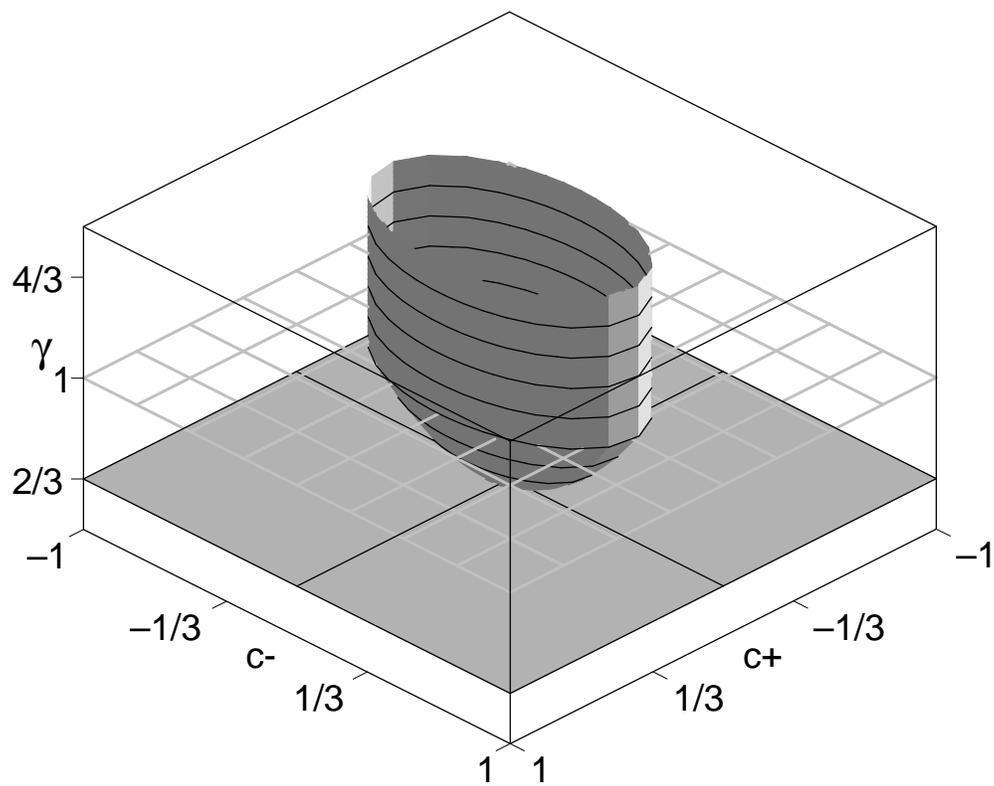}}
\caption{Stability zones of equilibrium points for the case
$\Omega_{\mu}=0$. Below the solid plane the point $1'$ is stable, between
solid and dashed planes outside the compact region the point $2'$ is stable,
inside the compact region the point $3'$ is stable, above the dashed plane
outside the compact region the point $4'$ is stable.
}
\end{figure}

 The part $\gamma>1$ in Fig.2  is new (in Fig.1 this part
would correspond to the unphysical region $\gamma>2$). In this part of the space
$(\gamma, c_+, c_-)$ the stability does not depend on $\gamma$. As a remarkable feature
we can note appearance of the Kasner solution as a future attractor
for $3c_+^2 + c_-^2 > 1/3$. For $3c_+^2 + c_-^2 < 1/3$
the point $3'$ is stable.

As in the previous section, using the expression
\begin{equation}
\Omega_{\lambda}'=[(6-6\gamma) \Sigma^2  + 2(2-3\gamma)U]\Omega_{\lambda}
\end{equation}
we can conclude that all the models
with $0<\Omega_{\lambda}<1$ and $\gamma \in (0,2/3]$ isotropize.

Anisotropic brane attractors may naturally appear in the inflationary
brane scenario
if the transition from nonstandard dynamics to standard one occurs
at some moment after the inflation.
During the nonstandard brane phase the condition for inflation is
$\gamma < 1/3$ \cite{Varun}, after the end of the inflation the parameter
$\gamma$ grows to $4/3$ in typical models.
 In this case the brane Universe evolves
consecutively from one stable regime to another one according to a corresponding
$c_+$=const, $c_-$=const line in Fig.2. When the pressure becomes
positive ($\gamma>1$), the Universe finally reaches one of the two possible
attractors depending on the parameters $c_+, c_-$ and stays in this regime until
the matter density on the brane drops below the brane tension and 
the transition to the standard regime occurs.

Near the initial singularity the nonstandard matter on the brane dominates
and we can put the standard matter density $\Omega_{\mu}=0$. As we now consider
the backward time direction, the condition for the stability
of the equilibrium points
becomes $\lambda_i>0$. The results are follows.

For the Kasner set two nonzero eigenvalues are $\lambda_1 = 6 - 6\gamma$ and
$\lambda_2=1-6\Sigma_+c_+-2\sqrt{3}\Sigma_-c_-$. Four other eigenvalues are
equal to zero. This indicates that we have a four-dimensional set of equilibria.
The second nonzero eigenvalue is positive on some subset of the Kasner set
and this subset exists 
for arbitrary $(c_+, c_-)$. On the other hand, $\lambda_1$ is positive
only for $\gamma<1$. So, the condition for
stability of the Kasner solution in the past is $\gamma <1$ independently
on $(c_+, c_-)$.

In the opposite case $\gamma >1$ (a matter with a positive pressure)
a stable solution in the past direction is the isotropic Universe.
This result  appears only
in the brane scenario. In the case of ${\cal P}_{\mu\nu}=0$
the isotropic
character of a brane cosmological 
singularity in the presence of matter with a positive
pressure has been found earlier by several authors \cite{Varun, C-S1, Topor}.

\section{Conclusions}

We have investigated the Bianchi I cosmological dynamics on a brane in the
presence of a nonlocal anisotropic stress ${\cal P}_{\mu\nu}$
satisfying the Barrow-Maartens
ansatz (9) with ${\cal U} >0$. The dynamics appears to
be more complicated in comparison with the previously known
case of ${\cal P}_{\mu\nu}=0$. However, we can mention two important regimes
not modified by the anisotropic stress of the form studied in this paper:

\begin{itemize}
\item If $0 < \gamma \le 4/3$ the anisotropic stress does not prevent the
Bianchi I model from isotropization. Using the existence of a monotonic
function (19) we can prove that all such models isotropize.

\item  The anisotropic stress does not alter the past asymptotic state
which is the isotropic Universe for $\gamma >1$ and the Kasner solution for $\gamma < 1$.
\end{itemize}

We also describe the set of future asymptotic states in the case $\gamma >4/3$
and possible transient brane dynamics which can exist
before the linear energy-momentum
contribution in the effective Einstein equations becomes important.

An important problem for the future development is to consider the ${\cal U} < 0$
case.

\section*{Acknowledgments}

The work was partially supported by RFBR grants Ns. 00-15-96699
and 02-02-16817.
AT is grateful to Roy Maartens and Alexei Starobinsky for
useful discussions.

\end{document}